\documentclass[12pt,preprint]{aastex}

\usepackage{emulateapj5}

\newcommand{\oh}{OH(1720~MHz)\ }
\newcommand{\candy}{G349.7+0.2}
\newcommand{\nh}{$N_{\rm H}$}
\newcommand{\hi}{H\,{\sc i}}

\newcommand{\e}[1]{$10^{#1}$}
\newcommand{\ee}[1]{$\times 10^{#1}$}

\newcommand{\cm}[1]{~cm$^{#1}$}
\newcommand{\cms}{~cm$^{-3}$\,s}
\newcommand{\kms}{~km\,s$^{-1}$}
\newcommand{\ergs}{~ergs\,cm$^{-2}$\,s$^{-1}$}
\newcommand{\erg}{~ergs\,s$^{-1}$}
\newcommand{\du}{\,$d_{22}$}

\newcommand{\msol}{$M_{\odot}$}

\begin{document}

\title{{\em Chandra} detection of ejecta in the small-diameter 
supernova remnant G349.7+0.2}

\author{J. S. Lazendic\altaffilmark{1}, P. O. Slane\altaffilmark{1}, 
J. P. Hughes\altaffilmark{2}, Y. Chen\altaffilmark{3} 
and T. M. Dame\altaffilmark{1}}

\altaffiltext{1}{Harvard-Smithsonian Center for Astrophysics, 60 Garden street,
Cambridge MA 02138}
\altaffiltext{2}{Department of Physics and Astronomy, Rutgers, The State
University of New Jersey, 136 Frelinghuysen Road, Piscataway, NJ 08854-8019}
\altaffiltext{3}{Department of Astronomy, Nanjing University, Nanjing 210093,
P. R. China}

\begin{abstract}

We present high-resolution X-ray observations  of the small-diameter
supernova remnant (SNR) \candy\ with the {\em Chandra X-ray
Observatory}.  The overall SNR spectrum can be described by
two  spectral components.  The soft component is in ionization
equilibrium and has a temperature of $\approx 0.8$~keV;  the hard
spectral component has a temperature of $\approx 1.4$~keV, an
ionization timescale of $\approx 5 \times 10^{11}$\cms\ and enhanced
abundances of Si. The spatially resolved  spectral modeling shows that
  S may also be enhanced, at least in some regions.  
The enhanced abundances clearly
point to the presence of an ejecta component in this remnant. Using 
 the available \hi\ and CO data towards \candy\ we derive a column 
density of  $\approx
7$\ee{22}\cm{-2} along the line of sight to the SNR, which is
consistent with  our X-ray data. The X-ray morphology of
\candy\ is strikingly similar to that at radio wavelengths --- an irregular shell
with a brighter eastern side --- which is consistent
with expansion in a medium with a large scale density gradient. The
remnant is known to be interacting with a molecular cloud (from the presence
of \oh masers), but this interaction is probably limited  to the
central portion of the SNR, as seen in SNR IC~443.  We found that \hi\
clouds are present in the SNR region, which supports
the notion that \candy\ belongs to a class of remnants  evolving in
the intercloud medium  \citep[such are IC~443 and W44;][]{chevalier99},
which is also responsible for the remnant's  morphology.  \candy\ does not
have the mixed-morphology found for other maser-emitting SNRs studied
to date in X-rays, but its morphology can be explained by a projection
model for mixed-morphology SNRs. We have identified a point source
close to the center of the SNR with a luminosity of $L_{X}
(0.5-10.0~{\rm keV})  \sim (0.2-2.3)\times 10^{34}$\du\erg,  which is
consistent with that of the compact central objects  found in a few
other Galactic SNRs.

\end{abstract}



\section{INTRODUCTION}

\candy\ is one of the three brightest radio supernova remnants  (SNRs)
 in our Galaxy  \citep{shaver85}, yet until recently very  little
 was  known about it. It is a small  diameter remnant ($\approx
 2\farcm 5$) classified as shell-type with a  brighter eastern SNR
 limb \citep{shaver85}. Early H\,{\sc i} observations indicated that
 \candy\ is located at a distance of $18.3\pm4.6$~kpc
 \citep{caswell75}.  \oh maser  emission  has been detected towards
 \candy\ at a radial velocity of around +16\kms, 
implying  that the SNR is interacting with dense molecular
 gas \citep{frail96}, and establishing a kinematic distance of
 22.4~kpc.  The  molecular complex
 associated with the remnant was  mapped in the CO 1--0 line
 \citep{reynoso01}, while near-infrared and millimeter-line
 observations have identified the shocked portion of the cloud
 located towards the center of the remnant
 \citep{lazendic03}. The shocked gas has temperature around 45~K and
 density around \e{5}\cm{-3}, which agrees well with conditions
 required for \oh maser production \citep{lockett99}.

In the X-ray band \candy\ was first detected in the {\em ASCA} Galactic
 plane survey \citep{yamauchi98}. Follow-up pointed observations  with
 {\em ASCA} showed that the remnant is one of the most luminous SNRs
 in the X-ray band \citep{slane02}, with a thermal spectrum dominated
 by strong K$\alpha$ lines from Si, S and
 Fe. The global spectrum was well fitted by both single equilibrium and
 nonequilibrium ionization models, with a plasma temperature $\approx
 1$~keV. The {\em ASCA} data indicated  an age of $\sim 2800$~yr, an
 unabsorbed X-ray  luminosity of  
 $L_X(0.5-10.0~{\rm keV}) \sim 2\times 10^{37}$\du\erg, 
 and an energy  release of  
 $\sim 5\times 10^{50}$~ergs in the supernova explosion.  The inferred
 abundances were consistent with solar values, and the large derived 
X-ray emitting mass ($\sim 160$\msol) showed the
 gas to be dominated by swept-up interstellar material.  These results
 are consistent with the SNR location near a molecular
 cloud. However, a detailed  morphological assessment and spatially
 resolved spectral analysis was not possible due to the poor spatial
 resolution of {\em ASCA}.  

In this paper we present high-resolution X-ray observations of \candy\
 with the {\em Chandra X-ray Observatory}.  We discuss the SNR
 morphology and plasma properties, compare X-ray data with available
 atomic and molecular data, and  compare the properties of \candy\ 
with those of similar  SNRs. 


\section{X-RAY DATA}

\candy\ was observed with the Advanced CCD Imaging Spectrometer (ACIS)
 on board the {\em Chandra X-ray Observatory} on  13 May 2002 (ObsID
 2785). The SNR was positioned on the ACIS-S3 chip, which has the best
 sensitivity and energy resolution in the soft X-ray bands.   Data
 were taken in full-frame timed-exposure mode with the standard
 integration time  of 3.2~s. 

 We used the contributed CXC software CORR\_TGAIN  \citep{Vikhlinin} 
to correct for time-dependence  of the ACIS gain \citep{grant03}. 
  After applying the corrections, the overall lightcurve was 
 examined for possible
 contamination  from a time-variable background. After these data
 reduction processes, the effective exposure was 56~ks from which
 48606  photons were observed for the whole remnant.

Data were further reduced using standard threads in the  {\em Chandra}
Interactive Analysis of Observations (CIAO) software package v.3.0.2, 
and Calibration Database (CALDB) v.2.26  which includes the
corrected FEF files for the S3 CCD used for making the ACIS response matrices.
We used weighted response matrices for spectral fitting 
to account for CTI effects across the CCD.
The low-energy (E $<$ 1~keV) quantum efficiency of the ACIS has been
degraded by molecular contamination on the optical blocking filter,
 and correction for this is integrated in the v.3.0.2. CIAO tools for 
 making auxiliary response files.  


\section{Data Analysis and Results}

\subsection{Spatial Analysis}

\begin{figure}
\centering
\caption{{\bf [see 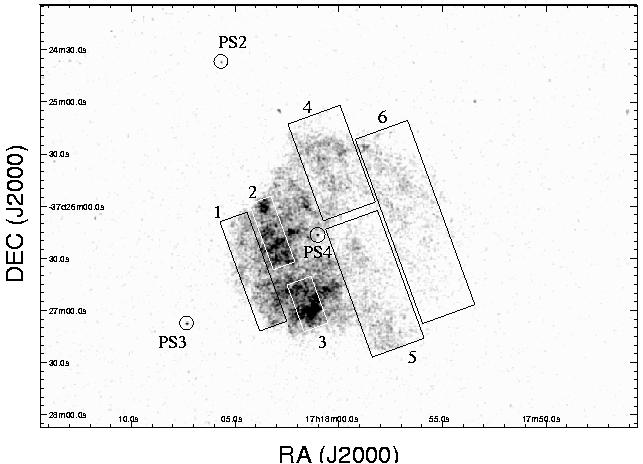]} The 1.0--8.0~keV band ACIS image of \candy. Three of five
(see Table~\ref{tab-psou}) point sources present in the field 
around \candy\ are marked with circles and letters PS. Six regions used for spectral 
analysis (each with $\approx 5000$ counts) are also marked.}
\label{fig-chandra}
\end{figure}

In Figure~\ref{fig-chandra} we show the exposure-corrected
 1.0--8.0~keV band ACIS image of \candy. The image has a  pixel size
 of $0\farcs 5$ and was smoothed with a Gaussian  filter with a FWHM
 of 1\arcsec.  The  emission is dominated by the eastern SNR shell,
 with the  peak emission located to the southeast. The bright region
 appears complex with several knots of emission. The western SNR shell
 is weak and diffuse,  with no noticeable structure. As shown later,
 this X-ray morphology is very similar to that in the radio band. 
 The images in different energy bands imply that the emission from the
  eastern-most part of the SNR shell may be softer with respect to
 the rest of the SNR.

Five unresolved sources have been detected in the field of \candy\
using  the CELLDETECT procedure in CIAO. Their properties  are
listed in Table~\ref{tab-psou}, and three sources are  marked with
circles in Figure~\ref{fig-chandra}. Source PS4 (CXOU J171801.0--372617) 
is the only source located, in projection, within
 the SNR shell.  All the sources have about the
same number of counts, except for source PS1. The positions of 
sources were compared with  the US Naval Observatory  Guide Star
Catalogue\footnote{Available at http://cadcwww.dao.nrc.ca/usno/} and
 the 2MASS All-Sky Point Source Catalogue\footnote{Available at
http://irsa.ipac.caltech.edu/}. For sources within 2\arcmin\ of 
the telescope axis, which is true for almost all of the sources 
 (Table~\ref{tab-psou}) the radius (90\% confidence) 
of the {\em Chandra} position uncertainty is 0\farcs 6\ (see {\em
Chandra} Proposer's Observatory Guide). 
Only source 1 has an optical/NIR counterpart (USNO 0525--27848790). The rest of the 
sources do not have counterparts within 5\arcsec\ of their position
down to 22~mag in $J$ band. 
Their spectra are hard, so they could be pulsars or background 
active galactic nuclei. Source 4 is of particular interest since it is
located within the SNR and we give it more consideration in the 
next section.

\subsection{Spectral Analysis}

\begin{figure}
\centering
\includegraphics[height=10cm]{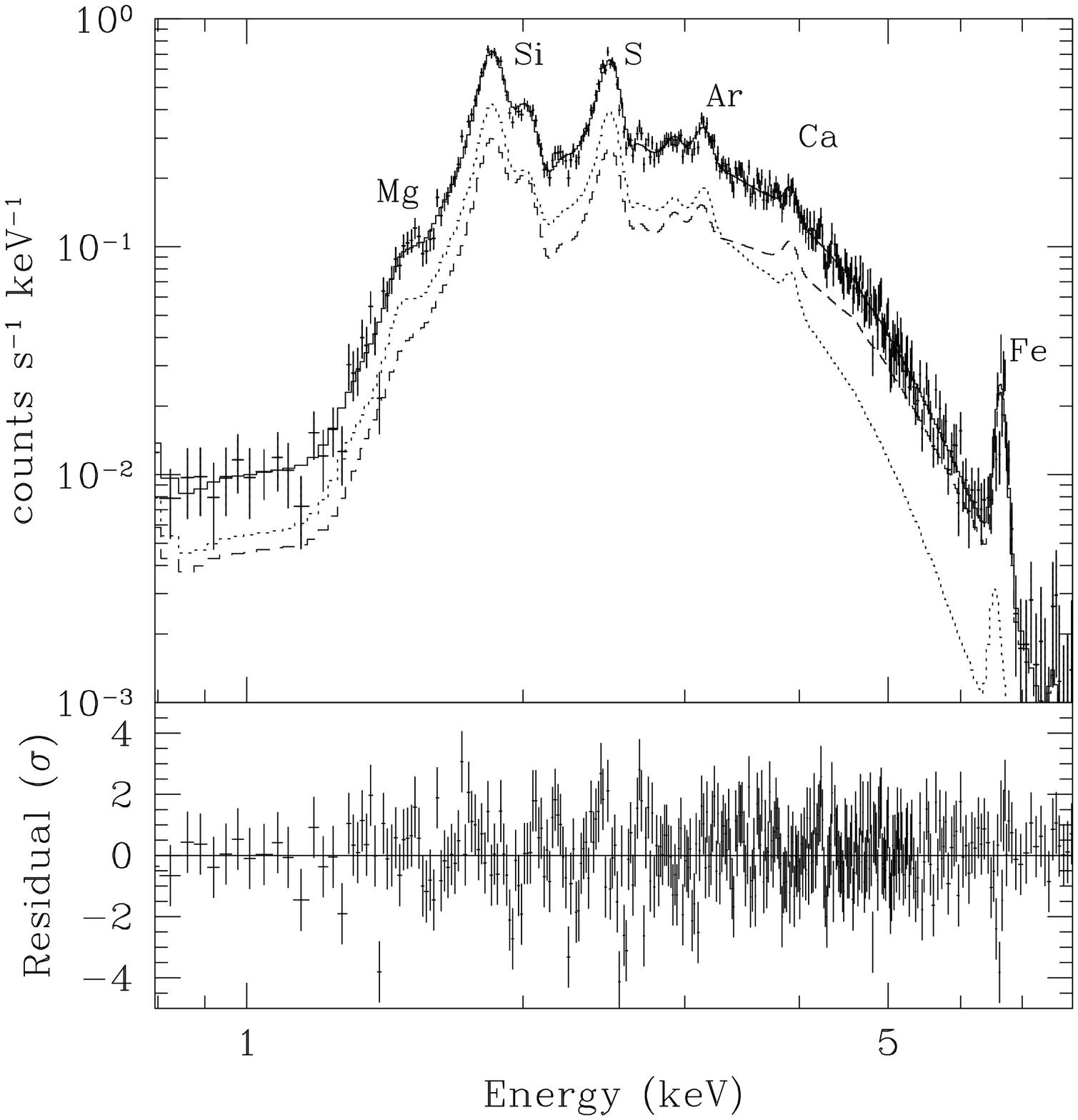}
\caption{ACIS spectrum extracted from the whole SNR, fitted with
two-component temperature (VRAYMOND+VPSHOCK+GAUSSIAN) model. The dotted
line represents the contribution of the soft component, the dashed
line represent the contribution of the hard component, and the solid
line is the sum of the both. Residuals from the model are plotted in
the lower panel.}
\label{fig-sp-whole}
\end{figure}

The spatially-averaged (global) spectrum from the entire remnant is shown  in
Figure~\ref{fig-sp-whole}.  A background spectrum was extracted from the
outer regions of the S3 chip. As observed previously by {\em ASCA},
the SNR spectrum is  dominated by K$\alpha$ line features from Si, S
and Fe.  In addition, the {\em Chandra} spectrum shows weaker
lines from Mg, Ar and Ca.  For spectral fitting we used models 
for an optically thin thermal plasma that has reached
collisional ionization equilibrium  \citep[VRAYMOND;][]{raymond77}
 and for a plasma that is still being ionized \citep[VPSHOCK;][]{borkowski01}.
 VPSHOCK is a plane-parallel time-dependent ionizing 
plasma model for which  the ionization timescale,
$\tau=n_e t$, describes  the progress of the plasma towards
equilibrium and can assume a range of values between the specified lower
(usually set to zero) and upper limits.   Since the version 
v.1.1 of the VPSHOCK model we used for the spectral fitting 
does not include the Ar line, we added 
a narrow Gaussian component at $\approx 3.1$~keV in our fits.

Single-component VRAYMOND or VPSHOCK models with 
solar abundances,  acceptable for
analysis of the overall {\em ASCA} spectrum from \candy,  were  not
able to give a satisfactory fit to the overall {\em Chandra} spectrum. 
 A good fit was obtained using a VPSHOCK model with enhanced Si
 abundance, and is significantly improved by adding an equilibrium
 thermal component with solar abundances; the best-fit parameters
 are listed in Table~\ref{tab-whole}. This result implies that the 
plasma in \candy\ is comprised of multiple
components, with a softer component having plasma temperature around
0.8~keV, and a harder component having plasma temperature around
1.4~keV.  We show the contribution of  both components to the
global spectrum in Figure~\ref{fig-sp-whole}. As expected, the harder
component (dashed line) dominates the spectrum above 3~keV, while
 the softer component (dotted line) dominates below that.
We find no strong nonthermal emission in \candy\
and estimate the power law component contribution to 
the observed flux from the whole SNR region to be 
less than 2.6\% for photon index values between 1.5 and 3 at the 3$\sigma$
confidence level.

The Si abundance in \candy\ is clearly enhanced; 
the $\chi^2$ value of the VPSHOCK+VRAY model drops by 46 
when one additional free parameter (the Si abundance) is added. Additional
elemental abundances were thawed one by one based on the inspection of the
residuals, but only the additional Si in the VPSHOCK 
(hard) component improved the fit significantly. 
This enhanced Si abundance implies that we
 have detected for the first time a  SN ejecta component in
this SNR. This also implies that the soft spectral component, which
has solar abundances,  is most likely related to the forward shock,
i.e.,  shocked interstellar material.
 The ionization timescales obtained for the two components are
consistent with this scenario. The forward shock is encountering interstellar
 material that is denser than that of the SNR interior, and will
therefore reach ionization equilibrium faster than shocked ejecta that
is expanding in the rarefied SNR interior.
The hydrogen column density obtained from the global {\em
Chandra} spectrum is consistent with that obtained from the {\em ASCA}
data \citep{slane02}. Such a large column density confirms a
 considerable absorption of the soft emission from the SNR and is
consistent with its relative distance. The X-ray luminosity derived
from the global SNR spectrum is $L_X(0.5-10.0~{\rm keV}) \approx 3.7 \times
10^{37}$\du\erg. We note that the value obtained for 
the plasma temperature from  the
global {\em ASCA} spectrum represents an intermediate value between the two
temperatures detected with {\em Chandra}.

The spectra from the whole SNR can therefore be fitted 
with two plasma components.  Could these two components be separate
 parts of a single evolving shock? To test this we also fitted the 
spectral regions
with a single planar, initially  unequilibrated shock
\citep[from][]{rakowski03}. In this model the temperatures of the
electrons and protons equilibrate downstream through Coulomb
collisions, providing a natural range in temperatures and ionization
timescales behind the shock. However, even in this model the best-fit
ionization timescales of the Si and S lines differ significantly
such that no single component model can accurately fit both line
complexes at once.

The high spatial resolution of the {\em Chandra} data allow us to
perform a spatially  resolved spectral analysis of \candy\ to search for
spectral variations across the SNR.  We selected 6 extraction
regions, shown in Figure~\ref{fig-chandra}, positioned along the
 SNR brightness gradient and each containing around 5000 counts.  
Spectra were re-grouped to include at least 25 counts per bin.  
 As for the global SNR spectrum, the spectrum used for background
subtraction was  extracted from the outer regions of the S3 chip.

\begin{figure}
\centering
\includegraphics[height=7cm]{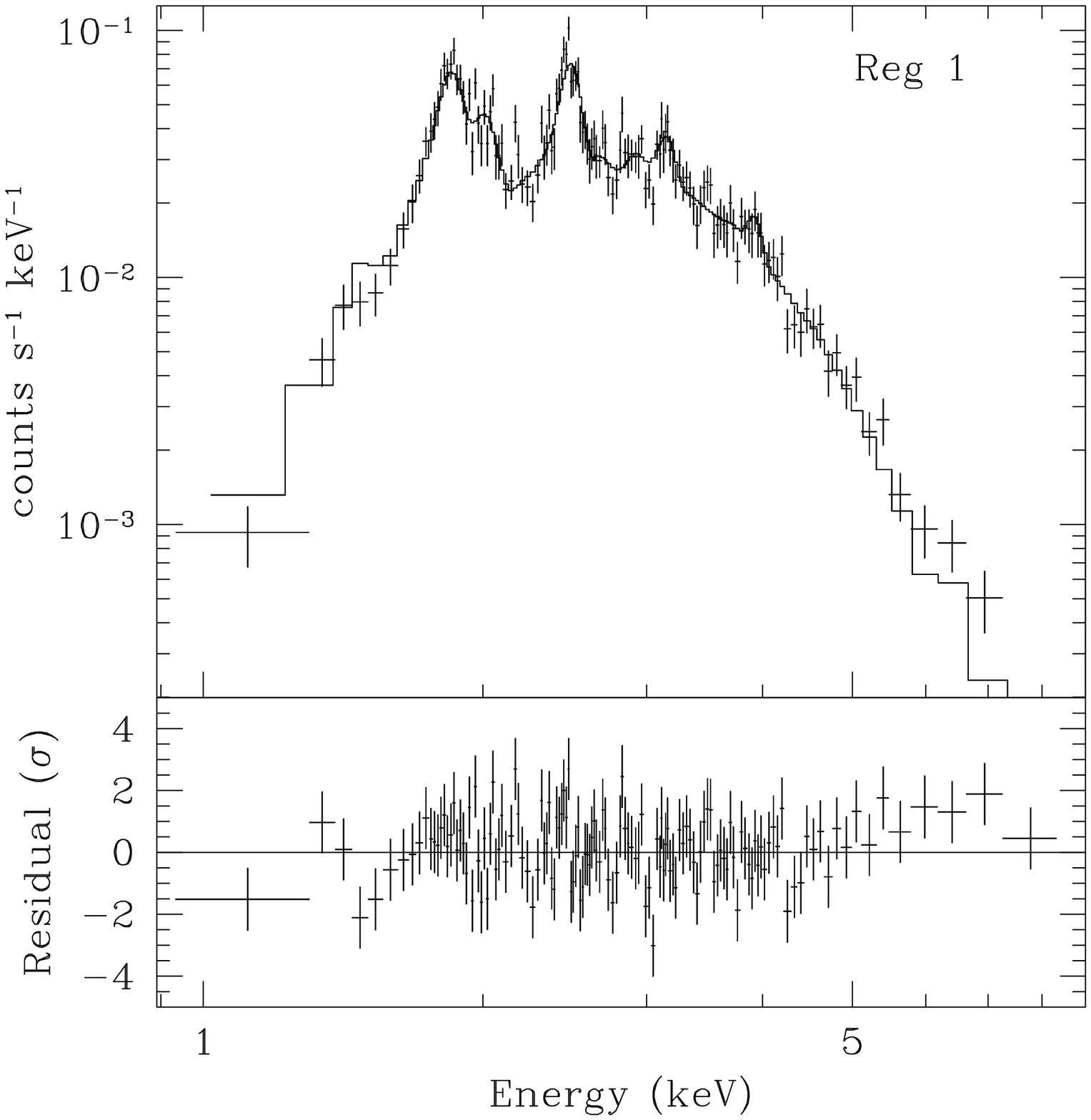}
\includegraphics[height=7cm]{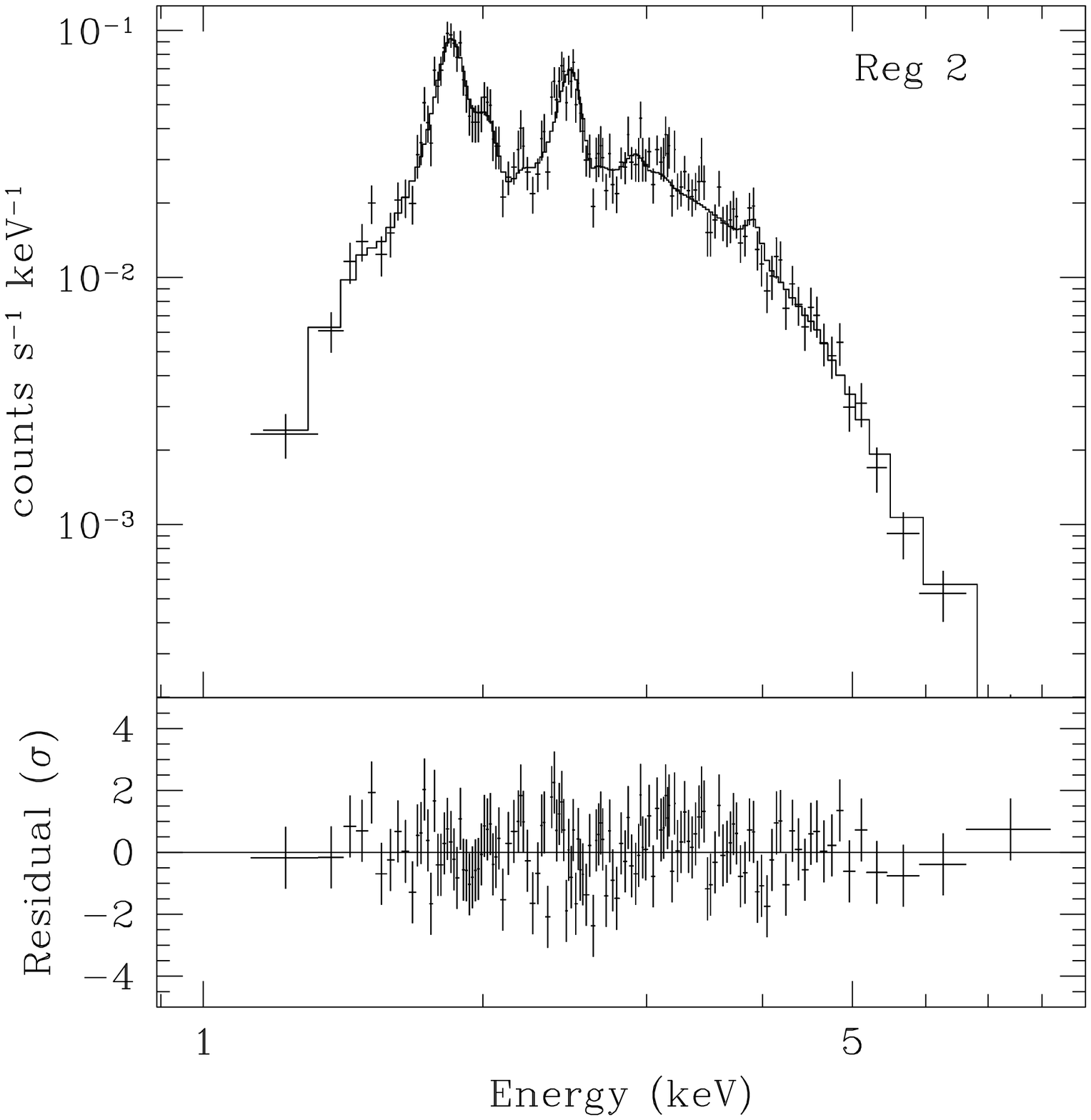}\\
\includegraphics[height=7cm]{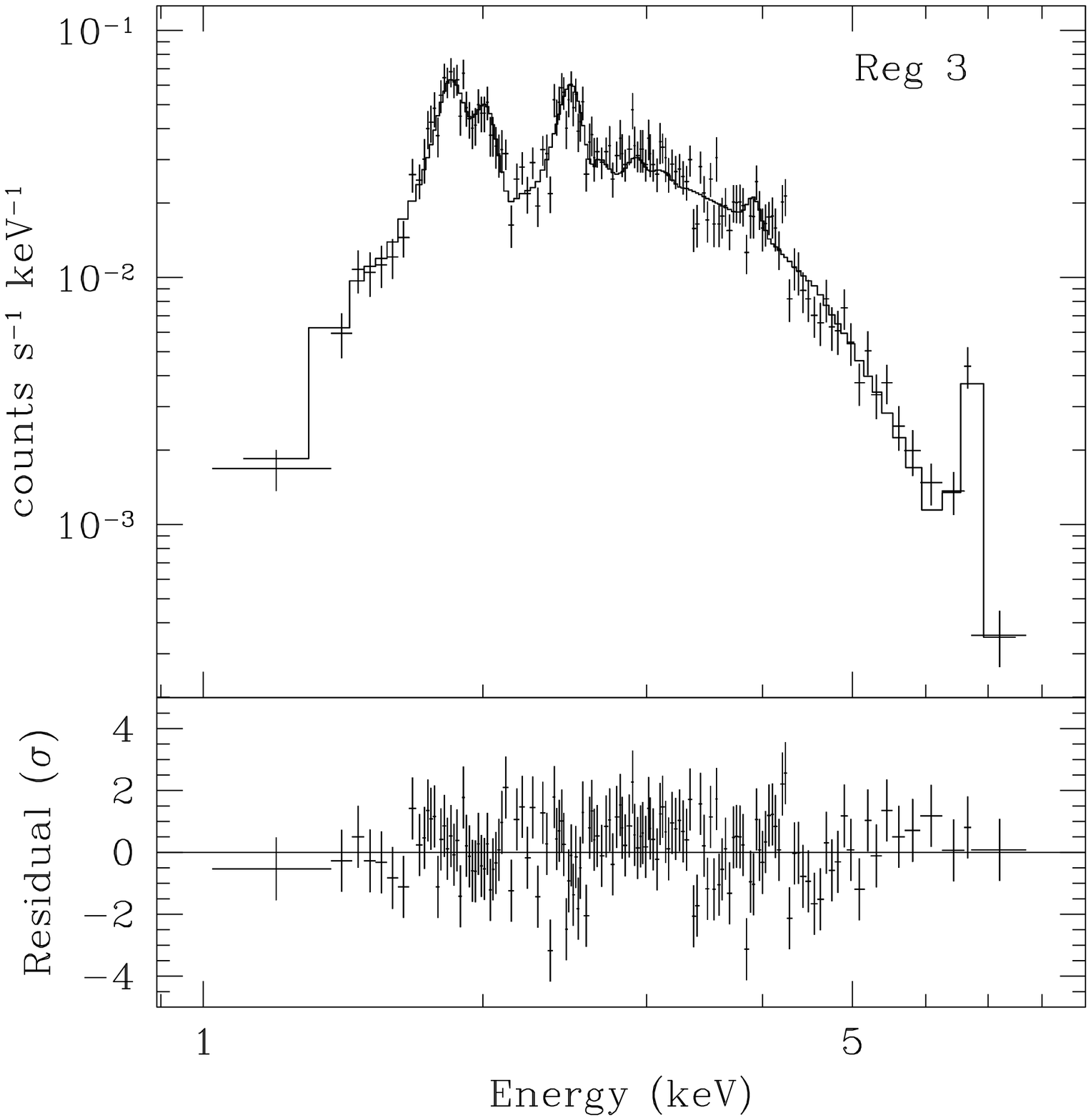}
\includegraphics[height=7cm]{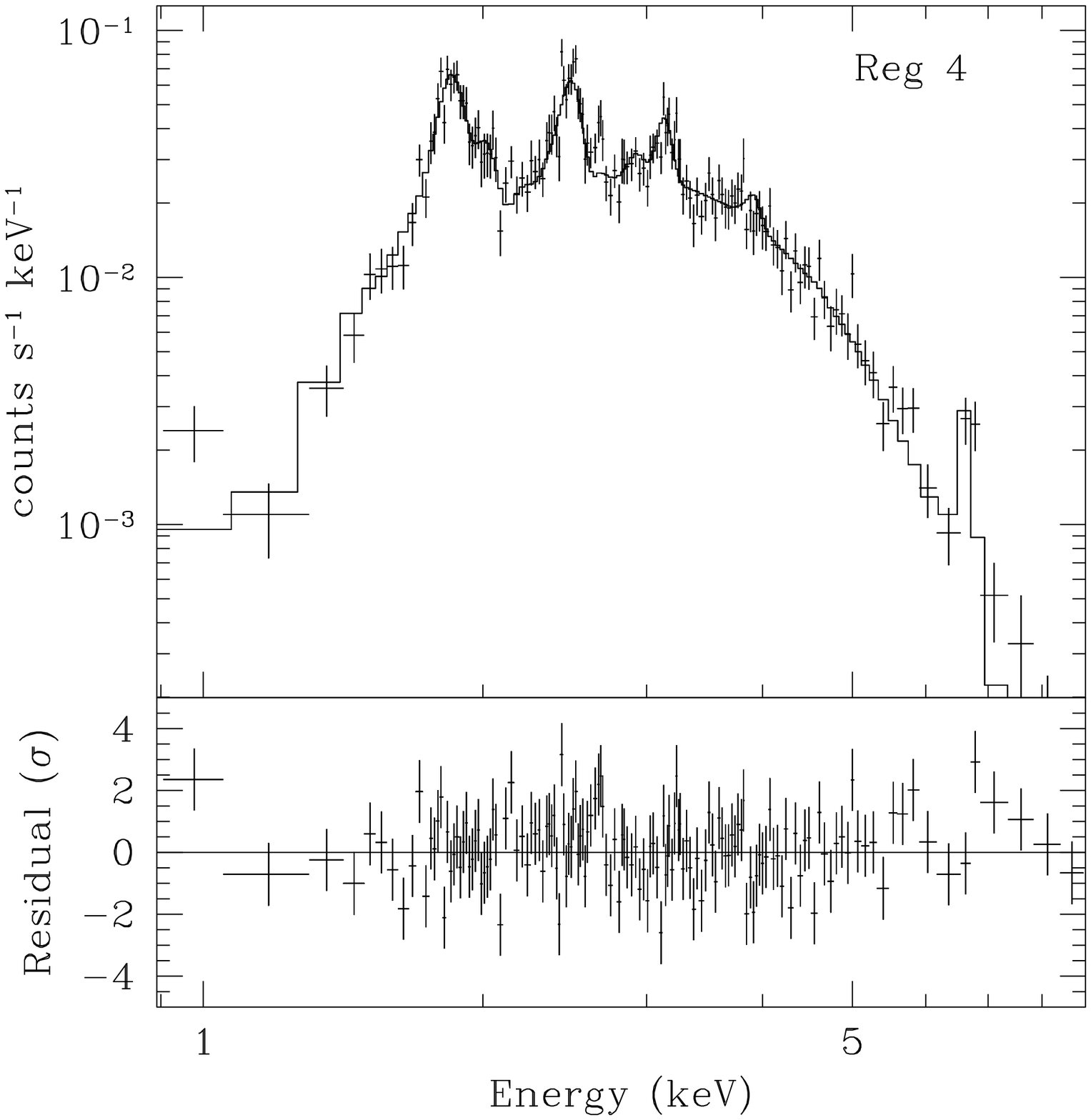}\\
\includegraphics[height=7cm]{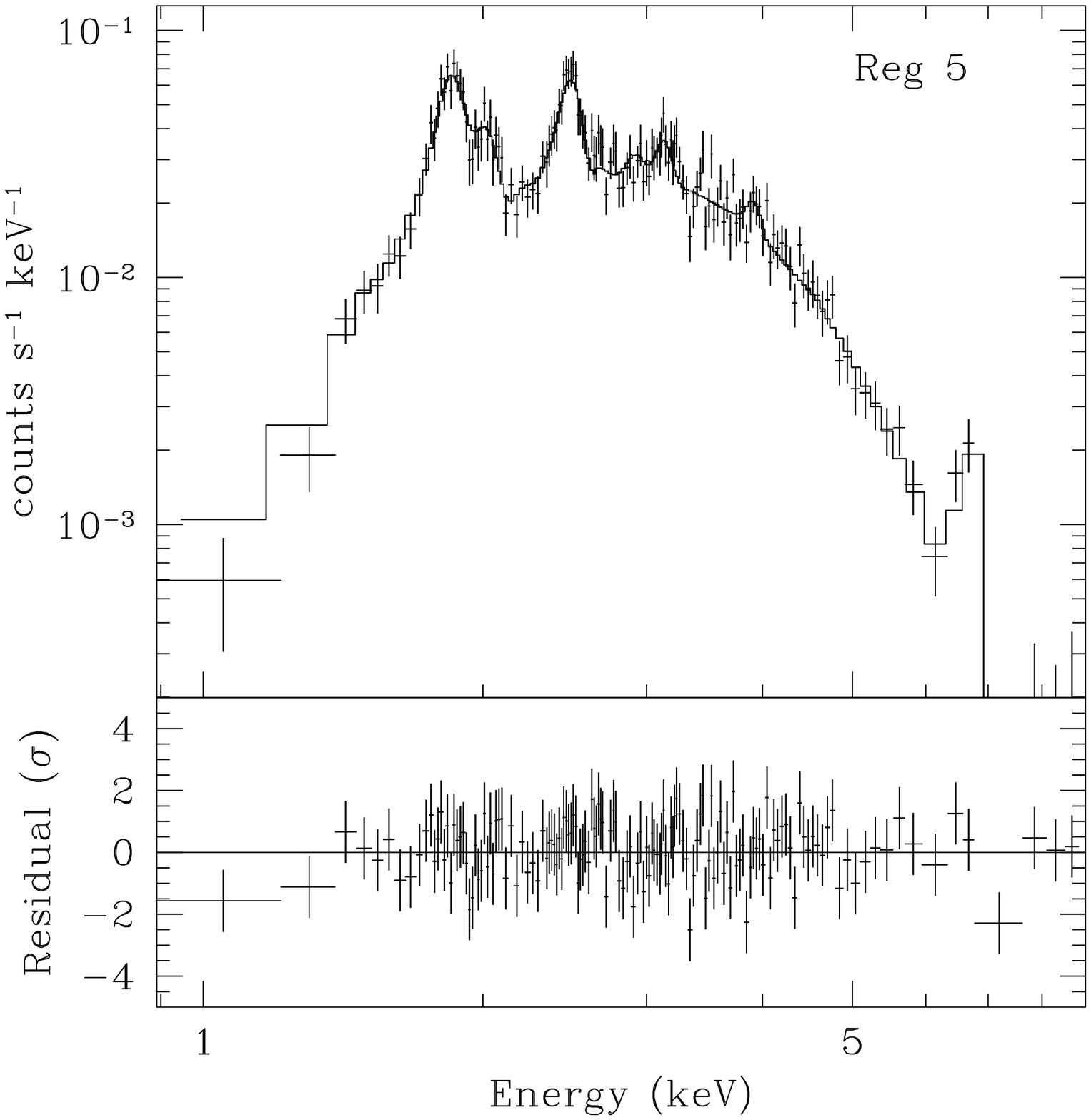}
\includegraphics[height=7cm]{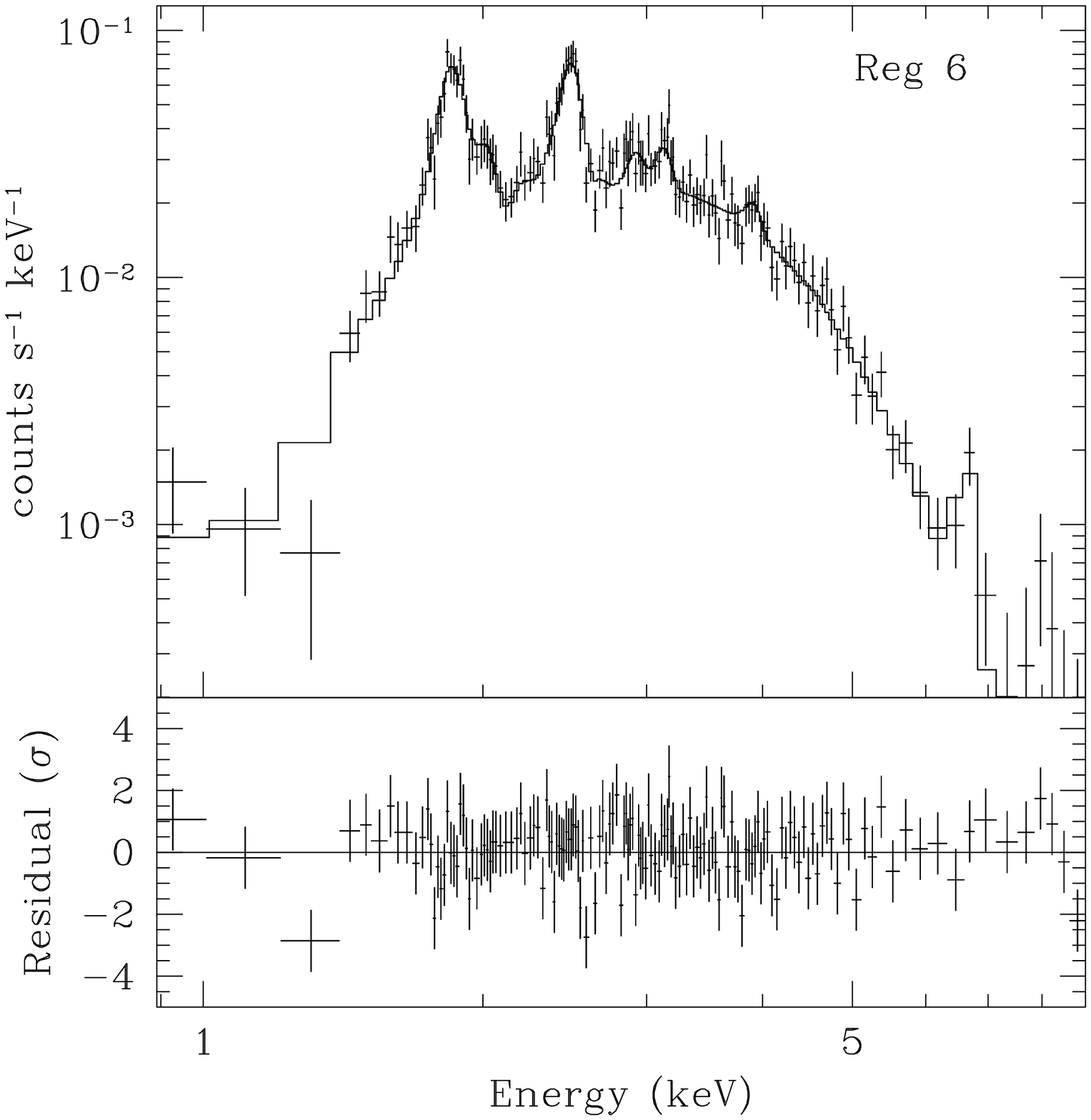}\\
\caption{ACIS spectra from the six SNR regions marked in
Figure~\ref{fig-chandra} and residuals from the best-fit models. 
Fits parameters are listed in Table~\ref{tab-fit}.}
\label{fig-spectra}
\end{figure}

\begin{figure}
\centering
\includegraphics[height=7cm]{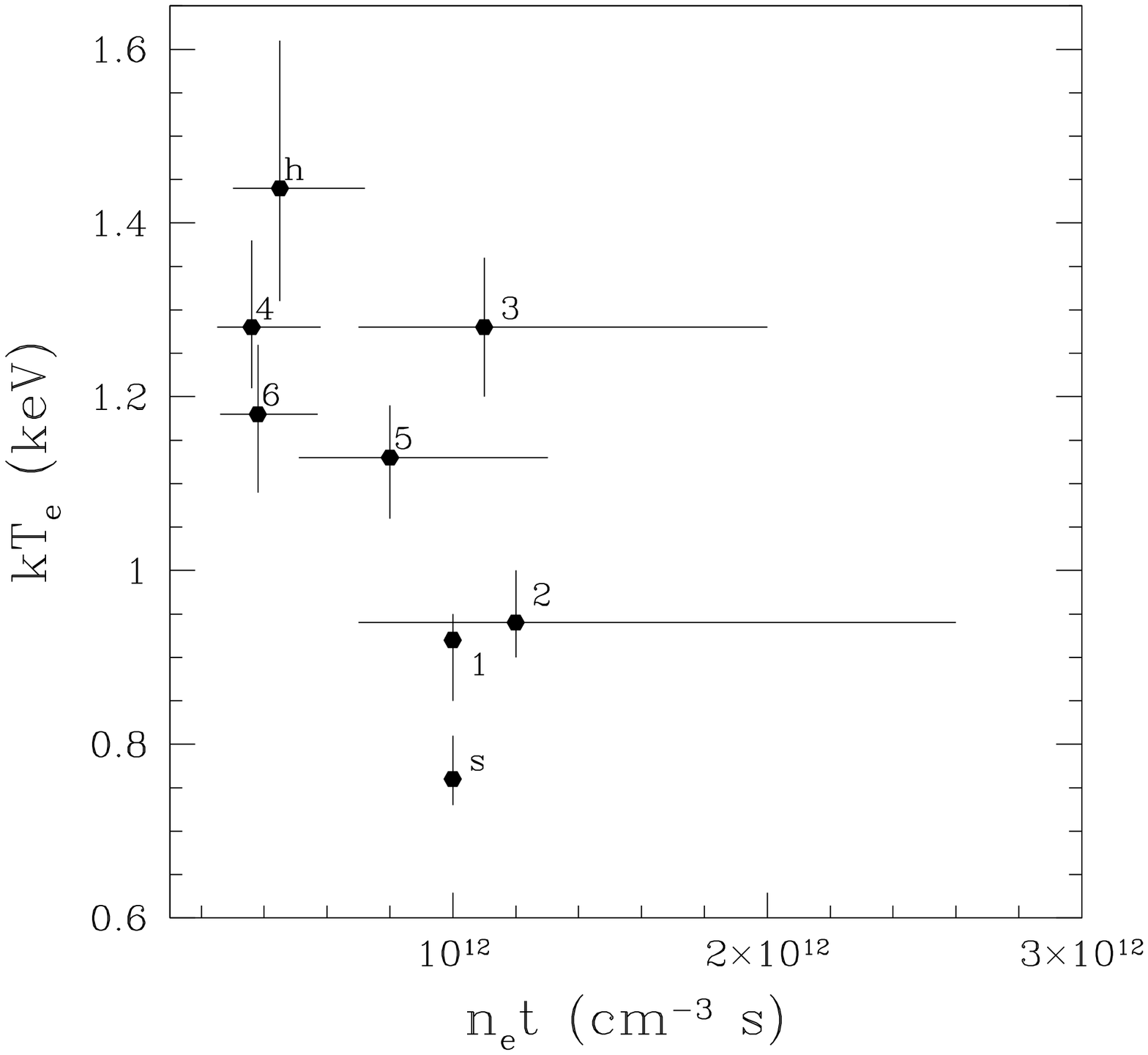}
\includegraphics[height=7cm]{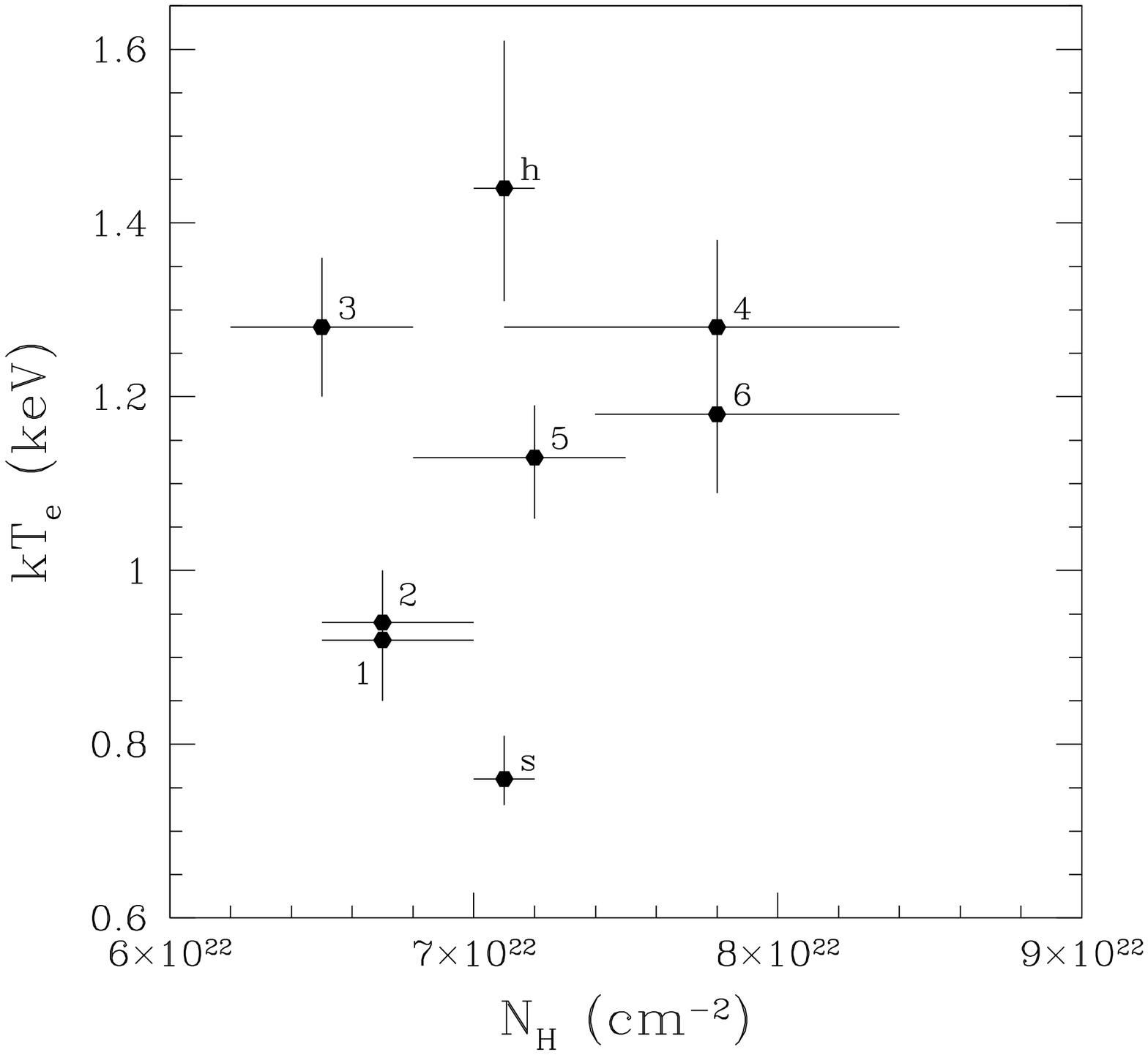}
\caption{The best-fit values and the 90\% confidence ranges 
of different spectral parameters measured in the six SNR regions (marked with
numbers) and in the whole SNR (s=soft component, h=hard component). 
({\em Left:}) Temperatures and ionization timescales.  ({\em Right}) 
Temperatures and column densities. The plasma temperature in the six
spectral regions is intermediate between the temperature measured from
the whole SNR, and the eastern-most regions (1 and 2)
have the plasma temperature lower than the other regions.}
\label{fig-temp}
\end{figure}

Unlike the overall SNR spectrum, spectra from the individual SNR
regions  can be well fitted with single-component thermal models whose
parameters vary from region to region; the fit
parameters are listed in Table~\ref{tab-fit} and the spectra are shown
in Figure~\ref{fig-spectra}.   All the regions, except the
 eastern-most region (region 1), favor the VPSHOCK model over the
VRAYMOND model; this implies that most of the plasma in  \candy\ has
not yet  reached ionization equilibrium and  is still evolving. 
Furthermore, all the
regions except the eastern-most region prefer the presence of enhanced
Si. Similarly, only the western-most region (region 6) prefers, 
in addition  to Si, an enhanced abundance of S. For
completeness, in Table~\ref{tab-fit} we list the values of Si 
and S abundances for all the regions. Plasma temperatures, 
 ionization timescales and column densities measured from the 
individual regions and the global SNR spectrum are plotted in
Figure~\ref{fig-temp}.  The temperatures derived from
individual regions have values intermediate to those of the 
soft and hard spectral components derived from the global
spectrum. This suggests that the individual regions do not 
have sufficient statistics to distinguish 
between the two plasma components detected in the global SNR spectrum, 
and represent an average of the two components summed along the line
of sight. The two eastern 
regions, regions 1 and 2, appear to have a lower plasma
temperature than the other regions, suggesting that the softer
component dominates in the eastern SNR region. Similarly, the
fact that these two regions do not show the  presence of Fe-K line
emission like the other regions suggests that plasma temperature in these two
regions is lower than elsewhere in the remnant. 
Figure~\ref{fig-temp} also implies that
 there may be a small variation in 
ionization timescale and column density across the SNR (e.g., regions 1, 2 and 3 versus regions 4 and 6).

\begin{figure}
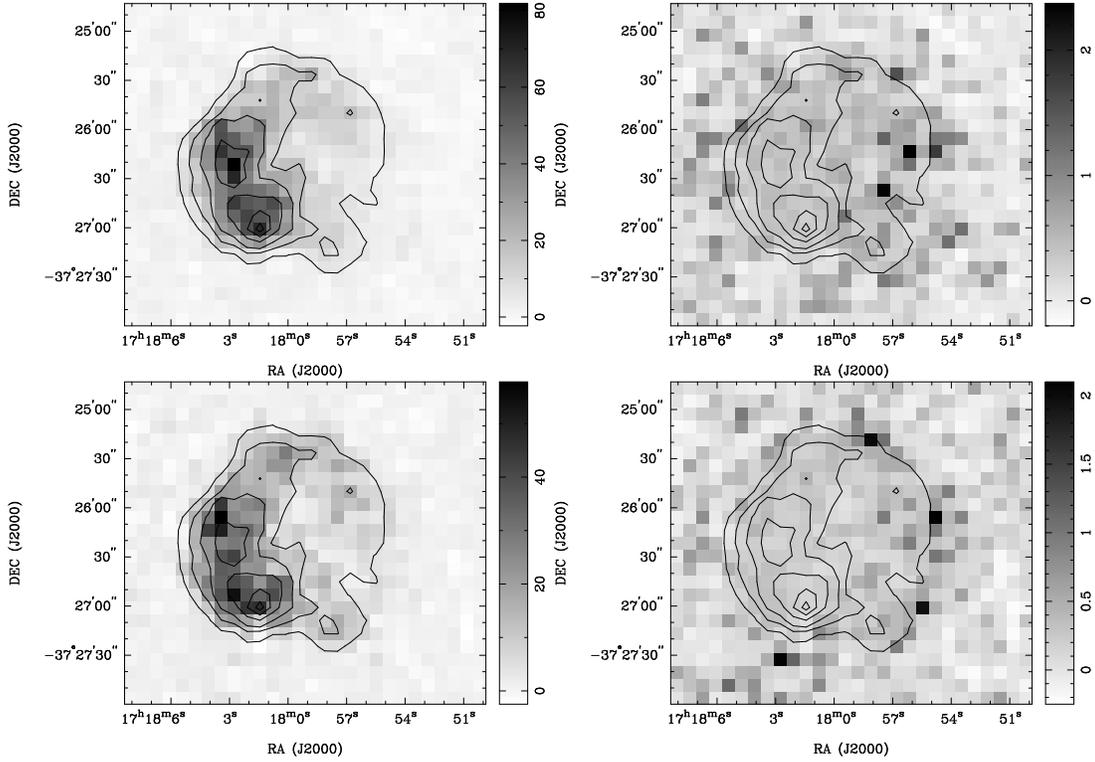

\centering
\includegraphics[height=5cm]{f5a.ps}
\includegraphics[height=5cm]{f5b.ps}
\includegraphics[height=5cm]{f5c.ps}
\includegraphics[height=5cm]{f5d.ps}
\caption{CS ({\em left}) and EW ({\em right}) images of \candy\ in Si
({\em top}) and S ({\em bottom}) line, binned by 16 pixels 
(i.e., 1 pixel=8\arcsec). 
Linear scale is in counts pixel$^{-1}$ for the CS images and 
in keV for the EW images. The
images are overlaid with contours from the 1--8~keV band image binned
in the same manner. Contour levels are: 70, 180,
300, 500, 700 and 900 counts~arcmin$^{-1}$.}
\label{fig-ew}
\end{figure}

To investigate the spatial distribution of Si and S line emission 
across the  remnant on a finer scale than that of the spectral
extraction regions, we produced  continuum-subtracted (CS) and
equivalent-width (EW) images  \citep[see
e.g.,][]{park03}, shown in Figure~\ref{fig-ew}.   
Narrow-band images used for deriving the CS and EW
images were filtered in energy according to Table~\ref{tab-ew}, and 
binned by 8\arcsec\ to provide
 a sufficient number of counts in the faint SNR regions. We found that
the CS images for Si and S  are quite similar to the whole band
image. The EW images, which represent line strength compared to the
continuum, do not show any spatial variation for the Si and S lines, 
 which appears inconsistent with the
results obtained from spectral modeling, at least for S.  However, 
 the EW of the emission lines does not depend just 
 on the total number of emitting atoms that are present in the plasma,
but also on the temperature and ionization state of the plasma. 
 Region 6, where an enhanced abundance of S is derived from the spectral fit,
 is the same region where the ionization timescale is low, 
which also makes the line-to-continuum ratio lower
 than expected for an equilibrium plasma. The net result is that the S EW
 is relatively uniform across the remnant. This is confirmed by
measuring EWs directly from spectra of the two representative SNR
regions. The derived EWs for the Si(S) line in regions 2 and 6 are 
$0.40\pm0.03$($0.24\pm0.02$)~keV and $0.36\pm0.03$($0.28\pm0.02$)~keV, respectively.

\subsection{SNR parameters}

To derive mean electron density $n$ for the two plasma components, 
we approximate the X-ray emitting volume $V$ as a
sphere 6.4~pc in radius, and denote $f_s$ and $f_h$ as filling factors for the soft and hard emission components, respectively. 
The emission measure for each component ($i=h,s$) is 
$EM_i \approx (n^2_i f_i V)/(4 \pi d^2)$. 
Assuming that the soft and hard  thermal components are 
roughly in pressure equilibrium ($n_s T_s \approx n_h T_h$), 
we can derive their relative filling factors:
$f_h/f_s=(EM_h/EM_s)/(n_h/n_s)^2=(EM_h/EM_s)/(T_s/T_h)^2$.  
Here we assume that the plasma temperature we measure represents 
the temperature of all species (electrons and ions), although the
X-ray fits only provide the electron temperature.
Substituting in the fitted values from Table~\ref{tab-whole}, we find that 
$f_h/f_s \approx 1.05$.  However, in the case that the whole SNR volume is
 not emitting X-rays, we set $f_V = f_s \approx f_h$ to account for the
unknown volume filling factor. The electron densities are then 
$n_s \ga 5.0~f_V^{-1/2}$\du$^{-1/2}$\cm{-3} and 
$n_h \ga 2.7~f_V^{-1/2}$\du$^{-1/2}$\cm{-3}.  The
X-ray emitting mass can be found from $M_x=1.4 n_H m_H f_V V$, which
gives $M_s \la 136~f_V$\msol\ mass in the soft component 
and $M_h \la 40~f_V$\msol\ mass in the hard component.

\citet{hnatyk99} showed that the global characteristics of a SNR
expanding into a density gradient will be similar to those of the
Sedov (adiabatic) SNR, because there is a mutual compensation of
 the emission deficit from the low density regions by the enhanced emission
from the high density regions.  The shock temperature is thus related
to the plasma temperature as in the  \citet{sedov59} model: $T_{sh}
= 0.78 T_X \approx 7 \times 10^6$~K, for $kT_X \approx 0.76$~keV 
(using the temperature of the soft spectral component which is most 
probably the forward shock component, as suggested previously). 
The corresponding shock velocity can be calculated from 
$v_{sh}=(16 k T_{sh} / 3 \mu m_H)^{1/2} \approx 710$\kms, 
where $\mu=0.604$ is the mean atomic weight.
The SNR age is proportional to the shock radius and velocity,  
$t=2 r_{sh} / 5 v_{sh}$.  We adopt the
SNR radius of 1\arcmin, for which its dimensional size is $r \approx
6.4$\du\,pc, and derive an SNR age of $t \approx
3500$\du~yr. This is in agreement with the upper limits derived
from the ionization timescales of the two components of  $t\le 4500$~yr. 
The explosion energy is approximately
$E= 4.6 \times 10^{-25} (n_s/4)
(r_{sh}^{5}/t^2) \sim 1.3 \times 10^{50}~f_V^{1/2}$\du$^{5/2}$~ergs, using the
value of $n_s$ derived above.  This is a slightly lower energy than 
the canonical value of \e{51}~ergs, although the uncertainty on the density
$n_s$ could modify this estimate somewhat.

\subsection{Point Source}

The X-ray emission from the point source CXOU J171801.0--372617 
 found inside the SNR shell is hard
(Table~\ref{tab-psou}), and there is a possibility that this source is
 a neutron star associated with the SNR. Indeed, in our {\em
 ASCA} study we suggested that \candy\ could be the 
remnant of a core-collapsed supernova  \citep{slane02}. Because the total number of
counts  from the point source is too low for spectral analysis, we 
estimated the unabsorbed flux using the count rate
($\approx 5.7$\ee{-4}~counts~s$^{-1}$), the effective area and 
response matrix files for ACIS-S.
Neutron stars are found with a range of photon index values, with
 classical young pulsars having values between 1.1--1.7
\citep[e.g.][]{chakrabarty01} and more exotic
objects like compact central objects (CCOs) and anomalous X-ray
pulsars (AXPs) having a softer photon index of $\sim 4$ and a blackbody
component with temperature between 0.2--0.4~keV \citep{pavlov02,pavlov03}.
Thus, assuming the same column density as for the SNR (7\ee{22}\cm{-2}) and a
power law photon index between 1.1 and 4.0 we obtained
 an unabsorbed integrated flux in the 0.5--10.0~keV range
 of $F_{X}^{0} \sim$(0.4-4.1)\ee{-13}\ergs, which at a distance of
22~kpc gives an X-ray luminosity of  $L_X \sim (0.2-2.3)$\ee{34}\erg.
This luminosity is lower than the values found from the 
other young pulsars, and is closer to the luminosities found from CCOs, 
 such as the one in SNR G347.3--0.5 \citep{lazendic-cco}.
There is no obvious compact source at the location of our point source 
in the radio image, but some difuse emission is present at 
 the 15~mJy\,beam$^{-1}$ level. Using the 
number of sources as a function of flux density from the {\em
 Chandra} Multi-wavelength Plane Survey \citep{grindlay03}, we derive
 a chance probability of $\approx0.2$ for finding an unrelated source
 with the X-ray brightness of $\sim$\e{-13}\ergs\ (assuming a 
power law photon index of 1.7) within the boundary 
of the SNR. Thus, the X-ray luminosity, and lack of an 
optical and radio counterpart at the
present sensitivity of the observations 
is consistent with a possibility that point
source in \candy\ is an associated  CCO. 


\section{Column density towards the SNR}

We also examined 21-cm \hi\ emission towards the \candy\ region 
from the Southern Galactic Plane Survey  (SGPS) data with angular 
resolution of $\approx 2\farcm 5$ \citep[e.g.,][]{griffiths01}. 
The 21~cm spectrum toward the remnant shows very strong absorption
at all negative velocities ($R < R_{\odot}$) and at positive velocities up to
+8\kms. This places the SNR slightly beyond the solar circle on the far
side. Adopting the molecular cloud velocity of +16\kms\ and 
using a flat rotation curve beyond the solar circle, the
kinematic distance is 23.0~kpc, as found by \citet{frail96}.

The Galactic center CO survey of \citep{bitran97}
covers this region with beamwidth (1/8\degr) sampling. Integrating the
nearest CO spectrum over the range $v = -220$ to $+17$\kms\ and using the
CO-to-mass conversion factor $X = 1.8 \times
10^{20}$\cm{-2}\,K$^{-1}$\kms\ \citep{dame01} we obtain:
\begin{displaymath}
2 \times N({\rm H}_2) = 4.6 \times 10^{22}~{\rm atoms~cm}^{-2}.
\end{displaymath}
The 21 cm velocity-integrated intensity varies very smoothly in
longitude and latitude through the position of the remnant, 
 and to obtain the atomic column density we used the average
21 cm spectrum over a small rectangle 
($27\arcmin \times 21\arcmin$) surrounding the remnant, 
but excluding a smaller region ($15\arcmin \times 9\arcmin$) of absorption
directly toward the remnant. A
lower limit on the atomic column density can be obtain by assuming the
21 cm emission is optically thin. This yields:
\begin{displaymath}
N({\rm H\,\scriptstyle I}) > 1.5 \times 10^{22}~{\rm cm}^{-2}.
\end{displaymath}
Adopting a standard \hi\ spin temperature of 140~K yields 
an atomic column density of
\begin{displaymath}
N({\rm H\,\scriptstyle I}) = 2.8 \times 10^{22}~{\rm cm}^{-2}.
\end{displaymath}
Therefore:
\begin{displaymath}
N({\rm total}) = N({\rm H\,\scriptstyle I})  + 2 \times N({\rm
H}_2) = 7.3 \times  10^{22}~{\rm cm}^{-2}, 
\end{displaymath}
which agrees well with the value derived from X-ray observations.


\section{DISCUSSION}

\subsection{SNR Morphology and Interaction with Ambient Medium}

\begin{figure}
\centering
\caption{ {\bf [see 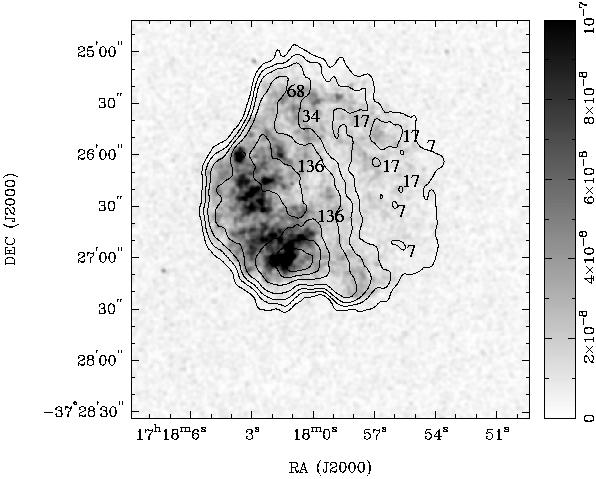]} The ACIS image overlaid with the Australia Telescope Compact
Array  18-cm contours from \citet{lazendic03}, with resolution of
$9\arcmin \times 5\arcmin$ (P.A.=$-1\degr$). Greyscale is in 
counts\,arcmin$^{-1}$\,s$^{-1}$. Most of the contour 
levels (in mJ\,beam$^{-1}$) are labeled on the image; the
peak two contours are at the 204 and 272~mJ\,beam$^{-1}$ level . The two
images show almost identical morphology.}
\label{fig-radio}
\end{figure}

The X-ray morphology of \candy\ revealed with the {\em Chandra}
observations is strikingly similar to that of the radio image 
\citep{lazendic03}, shown as contours in Figure~\ref{fig-radio}. Previous
studies in the radio band have suggested that the  structure of
\candy\ is consistent with two overlapping rings, with the fainter,
larger ring corresponding to the side where the SNR shock is
encountering a lower-density medium \citep{manchester87,reynoso01}.
This interpretation is consistent with theoretical modeling of SNR
expansion in a medium with a density gradient
\citep[e.g.,][]{tenorio85}. Since the SNR shock moves slowly in the
dense cloud and faster in the lower density medium, two half shells
with different radii are expected to be seen, with the shell radii
depending on the initial location of the explosion and the steepness
of the density gradient. In addition, \citet{hnatyk99}  performed 2-D
hydrodynamical modeling of the evolution and X-ray emission  of SNRs
expanding in a large-scale density gradient, as seen by an observer
under different viewing angles. Compared to their model,  the
morphology of \candy\ is broadly consistent with a large scale density
gradient seen by  an observer at an angle of $\approx 45\degr$.  

\begin{figure}
\centering
\caption{{\bf [see 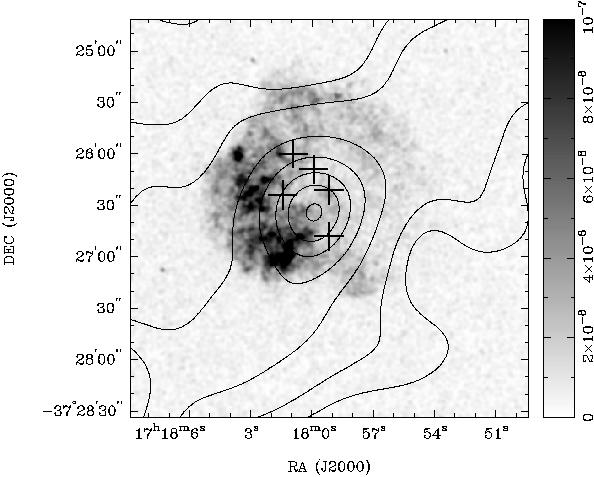, 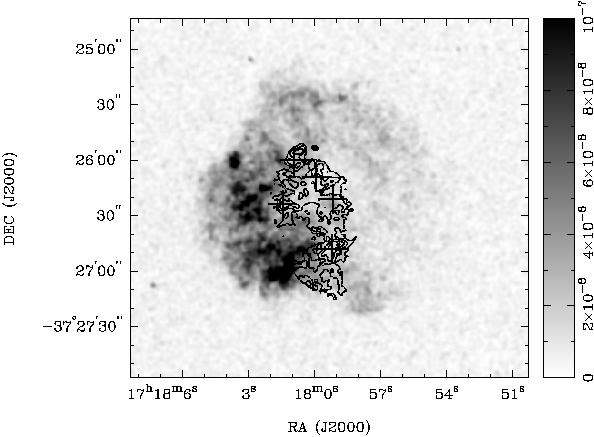]} Molecular line data towards \candy\ plotted as contours 
over the ACIS greyscale image. The crosses mark 
the \oh maser positions \citep{frail96}.
($Top:$) The velocity-integrated (+14 to +20\kms) CO 1--0 emission  
has 60\arcsec\ resolution \citep{reynoso01}. The contour levels are: 
4, 9, 20, 29, 35, 40 and 44 K\,km\,s$^{-1}$.  ($Bottom:$ ) The 
velocity-integrated 2.12~$\micron$ H$_2$ 1--0 S(1) line emission 
 has the spatial resolution of $\approx
1\arcsec$ \citep{lazendic03}. The contour levels are: 
 9, 17 and 35\ee{-5}~erg\,s$^{-1}$\,cm$^{-2}$\,sr$^{-1}$. }
\label{fig-mol}
\end{figure}

\begin{figure}
\epsscale{0.7}
\caption{{\bf [see 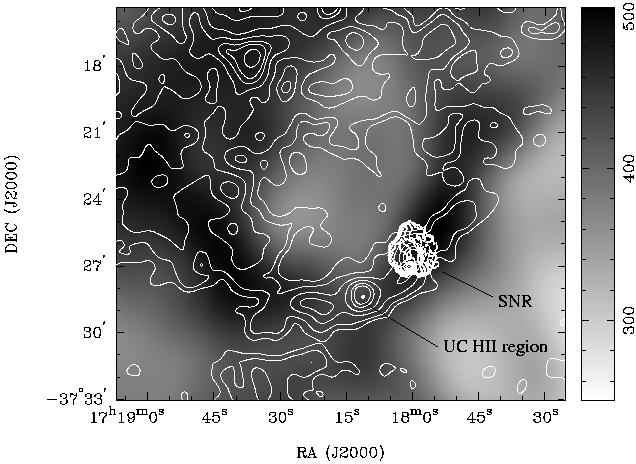]} 
Greyscale image of the velocity-integrated (+14 to +20\kms)
\hi\ emission towards \candy\ obtained from the SGPS 
\citep[e.g.,][]{griffiths01} with 2\farcm5\ resolution, overlaid with
CO integrated over the same velocity range \citep[white
contours;][]{reynoso01} with 1\arcmin resolution, and radio continuum
\citep[heavy white contours;][]{lazendic03}. The compact radio
continuum source is an ultra-compact H\,{\sc ii} region 
IRAS~17147--3725 \citep{bronfman96}. 
Greyscale is in K\,km\,s$^{-1}$.  The CO (white) contour
levels are 6, 10, 18, 26, 36, 41 and 46 K\,km\,s$^{-1}$.  The radio continuum
(heavy white) contour levels are 10, 17, 34, 68, 136, 205 and 274~mJy. \hi\
emission shows large scale correlation with CO emission.}
\label{fig-co+hi}
\end{figure}

\candy\ shows some morphological resemblance to another Galactic SNR
associated with \oh masers, IC 443. The two remnants are of similar
age and spatial size, and radio and X-ray morphology of IC~443 is
 also consistent with expansion into a large scale  density gradient
\citep{hnatyk98} produced by an \hi\ cloud on the
east and south-east side of the SNR \citep[e.g.,][]{braun86}. In
IC~443, an interaction with a CO cloud  is occurring in the central
region, where  shocked molecular gas and \oh maser emission is found
\citep[e.g.,][]{burton88,vand93,claussen97}.  In 
Figure~\ref{fig-mol}a we show the distribution of the molecular cloud
associated with \candy, as traced by the CO 1--0 
emission \citep{reynoso01}, superimposed on  the
{\em Chandra} image. The molecular cloud interaction 
with \candy\ also appears to be limited to
the central part of the remnant.  The location of \oh
masers \citep{frail96}, also shown in Figure~\ref{fig-mol}a, point to 
the on-going interaction between the CO cloud and the SNR.
 OH masers originate from shocked molecular clumps with density around
 \e{5}\cm{-3} and temperature of 50--125~K, which have encountered
the shock propagating transverse to the line of sight 
\citep{elitzur76,lockett99}. Shocked
H$_2$ 2.12~$\micron$ emission  has
also been detected at the locations of the OH masers \citep{lazendic03}. The
velocity-integrated H$_2$ 1--0 S(1) emission is shown in
Figure~\ref{fig-mol}b, superimposed on the {\em Chandra}
image. The H$_2$  line emission originates from the warm ($\sim 10^3$~K) 
gas just behind the part of the shock front expanding into the
molecular cloud.  The H$_2$ emission is also situated in the center of
the SNR where X-ray emission is very faint. It is possible that due to
high density in this region the SNR shock became radiative and we see
the  transmitted shock through the molecular cloud, that radiates
mostly in IR band; mid-IR and far-IR emission has also been
detected from this part of the SNR  \citep{lazendic03}. 
Thus, the density gradient responsible for the morphology of \candy\ is
probably produced by \hi\ clouds, as seen in IC~443.
Velocity-integrated (+14 to +20\kms) \hi\
emission is shown in Figure~\ref{fig-co+hi} overlaid with contours of CO
 integrated over the same velocity range and 
 the 18~cm radio image of the SNR. 
 The overall large scale morphology of the atomic gas is 
consistent with that of the molecular gas. The \hi\ gas follows the 
distribution of the large CO shell, with \hi\ emission peaks 
not always coinciding with the CO peaks, as is commonly found 
between atomic and molecular gas.  Thus, \hi\ material is present in 
the region towards \candy.

The density, size, and luminosity of \candy\ are similar to those proposed by
 \citet{chevalier99} for Galactic SNRs expanding into an intercloud
 medium, and to those of compact SNRs in other galaxies
 \citep{chevalier01}. These remnants are expanding into molecular
 clouds, but because of the clumpy structure of molecular clouds and
 the small filling factor of dense clumps, the remnant is mostly
 expanding into an intercloud medium with 
densities 5--25\cm{-3} \citep{chevalier99}. 
 Indeed, the molecular cloud associated with \candy\ has a density of
$\sim$\e{3}\cm{-3}, whereas densities determined from X-ray emitting
 plasma of $\sim 5$\cm{-3} are more characteristics of an intercloud
 \hi\ medium. This high density of ambient medium results 
in a rapid SNR evolution and a large amount of
 swept-up material producing very high luminosity ($\sim 10^{37}$\erg)
 in a short period
 just before the SNR  becomes radiative and starts to cool
 rapidly. Such remnants can be recognized by a high radio surface
 brightness \citep[e.g., IC~443 and W44;][]{chevalier99} or a
 strong IR  flux \citep{chevalier01}.  The X-ray observations of
 \candy\ imply that this SNR is still dynamically young enough to
 produce a significant X-ray flux, in  addition to strong radio
 flux. The radiative part of the shell should be accompanied by a \hi\
 shell.  However, high-resolution observations towards
 \candy\ are needed to investigate the association between the \hi\ emission
 and the SNR shock.

\subsection{Detection of Ejecta and the Two-Component SNR Plasma}

The solar abundances and large X-ray emitting mass derived from {\em
ASCA} observations imply that the X-ray emitting gas in  \candy\ is
dominated by swept-up interstellar material \citep{slane02}. The
 implication is that, similar to the case of the SNR N132D in  the
Large Magellanic Cloud \citep[e.g.][]{hughes87},  the SN ejecta 
component, expected from such a young SNR, is largely 
hidden by the glow of massive amounts of
swept-up material.  In the case of N132D,  optical studies  showed
clear evidence of fast-moving ejecta synthesized in the collapse of a
massive star. In X-rays, there was initially no detection of ejecta,
but high resolution spectral studies indicated enhanced levels of
oxygen \citep{hwang93}, and the presence of  an additional higher temperature
 spectral component which could be related to the ejecta
\citep{favata97,behar01}.  With the {\em Chandra} observations, we
clearly  detect enhanced abundances of Si and possibly of S in \candy, which
provides the first evidence for the presence of ejecta in this remnant.
Similar evidence for the presence of ejecta in two middle-aged Large
Magellanic Cloud SNRs, 0548--70.4 and 0534-69.9, with substantial
amount of swept-up material ($\sim 50$\msol) has been presented by 
\citet{hendrick03}, who derive enhanced Fe abundance based on the {\em
Chandra} spectra of these remnants.

The detection of the enhanced
abundances only in the harder spectral component suggests that this component
 is associated with the SN ejecta, which means that the softer
spectral component is related to the forward shock and the swept-up
material. If the ambient medium into 
which the SNR is expanding  has much lower density  than that of
the  ejecta, the forward shock will be heated to a significantly
higher temperature than the ejecta; this is the case in the young SNR
Cas~A \citep[e.g.,][]{willingale02}. However, \candy\ is expanding
into a dense medium and it is possible that the forward shock is
cooler than the ejecta. The assumed density gradient in the ambient
medium could yield enhanced S in just the western-most
region, as hinted at in the data; the amount of swept-up material 
is lowest there because of
the low ambient density, which enables the ejecta component to have
 a larger filling factor relative to the eastern regions where
 the ambient density is higher. Indeed, the fits to all the other regions (1
through 5) could accommodate a contribution from the plasma component of
 region 6 with an emission measure an order of
magnitude lower than the components listed in Table~3.

The lower limit on the fractional mass of ejecta can be inferred from 
$M_{ejecta}/M_{total} \ge (a-1)* M_{solar}/M_{total}$, 
where $a$ is the measured  abundance, and $M_{solar}/M_{total}$ 
is the standard solar mass fraction; for
Si and S  the mass fractions are 6.96\ee{-4} and 4.79\ee{-4}, respectively
\citep{grevesse98}.   Using the measured abundances from region 6, 
we derive $M_{\rm Si}/M_{total}\ge 9.74$\ee{-4} and 
$M_{\rm S}/M_{total}\ge 2.39$\ee{-4}. The observed emission measure in
region 6 implies a gas density of $n_H\approx 2.3$\cm{-3}, which for
 an emission volume of $\approx 1.4$\ee{58}\cm{3} ($32\arcsec \times
113\arcsec \times 144\arcsec$) gives a total mass in the region 
of $\approx 34~f_V$\msol.  The mass of the observed Si and S ejecta 
 is thus  $\ge 0.033~f_V$\msol\ and $\ge 0.008~f_V$\msol, respectively.
 This is a significant fraction of the amount produced by 13--15\msol\
stars \citep[Si: 0.047--0.07\msol, S: 0.026--0.023\msol;][]{thielemann96}.

\subsection{Implications for Mixed-Morphology SNRs}

\oh maser emission such as that associated with \candy\ and IC~443, 
 has been found in about 10\% of the Galactic SNRs  
\citep[e.g.][]{green97,wardle02}.  Seven 
of these maser-emitting \citep[ME;][]{yusef03} SNRs have been studied 
in detail in X-rays, and all of them are found
to be of the mixed-morphology \citep[MM;][]{rho98} class of  
SNRs. MM remnants, also known
as thermal composites,  are  identified by shell emission in the
radio band, and  centrally brightened thermal emission in the X-ray
band with little or no shell brightening.  The evolutionary
characteristics which lead to mixed-morphology X-ray properties in
SNRs are not well understood, and different models have been suggested
to explain individual cases.  One such model invokes evaporation of
clouds which are left relatively  intact after the passage of the SNR
blast wave and are evaporating in the SNR interior \citep{white91}.  
Another model explains the emission for the
mixed morphology SNR W44 \citep{cox99} as due to the effect of thermal
conduction in the remnant interior. Another 
model for mixed morphology SNRs invokes shell-like SNR
evolution at the edge of a molecular cloud, in which the ambient 
density gradient does not lie in the projection plane \citep{petruk01}.
 
A possible correlation between ME and MM SNRs has been suggested,
because both types of SNRs appear to be associated with molecular
clouds  \citep{green97,rho98}. The fact that some MM SNRs
do not have OH masers is not unexpected because very specific physical conditions
 are required for the maser excitation. However, it has been 
 suggested that all ME SNRs
 could be of the MM class \citep{rho98,yusef03}.  
 \candy\ is then an example of a ME SNR which does not exhibit 
 MM morphology. Its morphology, which is not centrally peaked, 
 can be explained by the projection effect
 model used to explain MM SNRs as well \citep{petruk01}. 
 As shown in previous sections, we  are
 likely seeing the SNR expanding into a density gradient of 
an angle of $\approx 45\degr$. If  we were to
 observe  the  expansion along the direction of the density gradient, we
would see centrally peaked X-ray emission rather than two overlapping
shells. Another example is the only ME SNR in the Large
Magellanic Cloud, N49 \citep{brogan04}. This remnant also has 
 an irregular shell morphology in both radio and X-rays, with  
 a bright region to the south-east \citep[e.g.,][]{park03}, which can similarly
be explained as resulting from expansion in a density gradient
produced by the presence of a molecular cloud, as mapped by \citet{banas01}.  
Thus, if ME and MM SNRs are the same class of the SNRs and
there is a single mechanism to explain their appearance and spectral
characteristics, the observations of \candy\ would support 
the projection model. 
Identifying more cases of SNRs expanding into density gradients seen
under different angles might provide further support for the projection
model and contribute towards understanding the nature of MM SNRs.


\section{CONCLUSIONS}

We present  high-resolution {\em Chandra} observations of a
small-diameter  SNR, \candy. Its X-ray morphology is almost identical
to that at radio wavelengths, and  is consistent with expansion into a
density gradient viewed at an angle of  $\approx 45\degr$. We
investigated the atomic and molecular material in the vicinity of  the
SNR,  and  suggest that the SNR belongs to the class of SNRs
expanding into an intercloud medium discussed by Chevalier (1999). 
 The interaction with the
molecular cloud is probably  limited to the central portion of the
SNR, where the SNR shock became radiative and the shocked material
radiates mostly in the IR band \citep{lazendic03}. \candy\ belongs to
a class of maser-emitting SNRs which have been suggested  to be related to the
mixed-morphology class of SNRs.  \candy\ does  not exhibit
mixed-morphology, but its brightness distribution 
 can be  explained with the projection model
for mixed-morphology SNRs.

The spectrum from the whole SNR can be fitted with
two thermal components, produced by distinct swept-up and ejecta
plasma contribution. The softer component has a plasma temperature
$\approx 0.8$~keV and is in ionization equilibrium, while the harder
component has plasma temperature $\approx 1.4$~keV, ionization  
timescale  $\approx 4 \times 10^{11}$\cms\ and enhanced abundances of
Si. The slightly enhanced S in just the western-most
part of the SNR could, if correct, be explained by a higher filling factor of the
ejecta component relative to the swept-up component because of the
lower density of the ambient medium in that region. The column density
 to the SNR of $\approx 7$\ee{22}\cm{-2} 
derived from X-ray data agrees well with the value
derived from \hi\ and CO data.

We find a point source with a small number of counts ($\approx 30$)
close, in projection, to the SNR center. Using the column density value found towards
the SNR and the observed range of photon index values from neutron stars,
we estimate the luminosity of the point source in the 0.5--10.0~keV band
to be $\sim (0.2-2.3)\times 10^{34}$\du\erg. This luminosity and the lack of
an optical and radio counterpart suggests that the X-ray point source 
 could be the compact object formed by the explosion that also
produced in \candy.


\acknowledgements{}

We thank Cara Rakowski for use of her shock model, 
John Raymond, Oleh Petruk and  Sangwook Park
for helpful discussions,  Estela Reynoso for the CO image and Bryan
Gaensler for the \hi\ data.   This work was supported in part from
{\em Chandra} grants GO2-3080A (JSL) and GO2-3080B (JPH), 
and NASA contract NAS 8-39073 (POS).


\clearpage

\begin{deluxetable}{llcccccc}
\tabletypesize{\scriptsize}
\tablecaption{Properties of Serendipitous Sources in the field of 
\candy.\label{tab-psou}}
\tablewidth{0pt}
\tablehead{
\colhead{Source}  & \colhead{Name} & \colhead{R.A.}   & \colhead{Decl.} &
\colhead{ACIS}  & \colhead{Pointing}& \colhead{Hardness} &
\colhead{Optical/IR} \\
\colhead{ }  & \colhead{ }  & \colhead{ }   & \colhead{ } &
\colhead{counts} & \colhead{offset} & \colhead{ratio\tablenotemark{a}}
& \colhead{counterpart}  \\
}
\startdata

1 & CXOU J171816.2--372556 & 17 18 16.19 & $-$37 25 55.84 & $72\pm9$ &
2\farcm 3 & $-0.6\pm0.5$ & yes, $J$(mag)=10 \\
2 & CXOU J171807.3--372707 & 17 18 07.32 & $-$37 27 07.66 & $30\pm6$ &
1\farcm 5 & $0.10\pm0.10$ & no   \\
3 & CXOU J171805.7--372437 & 17 18 05.67 & $-$37 24 37.41 & $26\pm5$ &
1\farcm 2 & $0.86\pm1.19$ & no  \\
4 & CXOU J171801.0--372617 & 17 18 01.00 & $-$37 26 16.62 & $32\pm6$ &
0\farcm 8 & $0.35\pm0.34$ & no  \\
5 & CXOU J171746.1--372506 & 17 17 46.10 & $-$37 25 05.90 & $28\pm6$ &
3\farcm 7 & $0.67\pm0.71$ & no  \\

\enddata
\tablenotetext{a}{The hardness ratio defined as $HR = (H-S)/(H+S)$, 
where $S$ is the number of counts from 1.0 to 2.5~keV and $H$ is 
the number of counts from 2.5 to 8~keV.}

\end{deluxetable}



\begin{deluxetable}{ll}
\tabletypesize{\scriptsize}
\tablecaption{Results from fitting VRAYMOND+VPSHOCK+GAUSS model to
 the whole SNR with the 90\% confidence ranges. 
 \label{tab-whole}}
\tablewidth{0pt}
\tablehead{
\colhead{Parameter}  & \colhead{Whole SNR}}
\startdata
\nh\ (\e{22}\cm{-2}) & 7.1$^{+0.1}_{-0.1}$ \\
\multicolumn{2}{l}{Soft component (CIE)\tablenotemark{a}} \\
$kT$ (keV)           & 0.76$^{+0.05}_{-0.03}$ \\
$EM$ (\e{59}\cm{-3}) & 9.9$^{+1.8}_{-2.3}$ \\
 \multicolumn{2}{l}{Hard component} \\
$kT$ (keV)           & 1.44$^{+0.17}_{-0.13}$ \\
$n_e t$ (\e{11}\cms) & 4.5$^{+2.7}_{-1.5}$ \\
Si                   & 2.1$^{+0.2}_{-0.3}$ \\
$EM$ (\e{59}\cm{-3}) & 2.9$^{+1.4}_{-0.5}$ \\
 \multicolumn{2}{l}{Gaussian line component} \\
$E$ (keV)            & 3.1 (frozen) \\
$dE$ (keV)           & \e{-5} (frozen) \\
$K$ (phot\cm{-2}\,s${-1}$) & 2.9$^{+2.3}_{-1.9}$\ee{-5} \\
$\delta \chi^2$/dof  & 1.22/339 \\
\multicolumn{2}{l}{Unabsorbed X-ray flux (0.5--10~keV)}\\
$F_{\rm total}$(\ergs)    & 6.5\ee{10} \\
$F_{\rm hard}$(\ergs)     & 1.7\ee{10} \\
\enddata
\tablenotetext{a}{collisional ionization equilibrium}
\end{deluxetable}



\begin{deluxetable}{lcccccc}
\tabletypesize{\scriptsize}
\tablecaption{Results from fitting VRAYMOND or VPSHOCK model to six spectral 
regions in SNR \candy\ marked in Figure~\ref{fig-chandra}  
with the 90\% confidence ranges.  \label{tab-fit}}
\tablewidth{0pt}
\tablehead{
\colhead{Parameter}  & \colhead{Reg 1} & \colhead{Reg 2}   
& \colhead{Reg 3} & \colhead{Reg 4}  & \colhead{Reg 5} & \colhead{Reg 6}
}
\startdata

\nh\ (\e{22}\cm{-2}) & 6.7$^{+0.3}_{-0.2}$
                     & 6.7$^{+0.3}_{-0.2}$
                     & 6.5$^{+0.3}_{-0.3}$
                     & 7.8$^{+0.6}_{-0.7}$ 
                     & 7.2$^{+0.3}_{-0.4}$ 
                     & 7.8$^{+0.6}_{-0.4}$ \\

$kT$ (keV)          & 0.92$^{+0.03}_{-0.07}$  
                    & 0.94$^{+0.06}_{-0.04}$ 
                    & 1.28$^{+0.08}_{-0.08}$ 
                    & 1.28$^{+0.10}_{-0.07}$
                    & 1.13$^{+0.06}_{-0.07}$
                    & 1.18$^{+0.08}_{-0.09}$ \\

Si                  & 1.4$^{+0.2}_{-0.3}$
                    & 1.6$^{+0.3}_{-0.2}$
                    & 1.7$^{+0.3}_{-0.3}$
                    & 1.6$^{+0.4}_{-0.2}$
                    & 1.6$^{+0.4}_{-0.2}$ 
                    & 2.4$^{+0.6}_{-0.6}$  \\

S                 & 1.2$^{+0.2}_{-0.2}$ 	
                  & 0.9$^{+0.2}_{-0.1}$
                  & 0.8$^{+0.1}_{-0.1}$
                  & 1.2$^{+0.3}_{-0.1}$
                  & 1.2$^{+0.2}_{-0.2}$
                  & 1.5$^{+0.2}_{-0.3}$ \\

$n_e t$ (\cms)         & CIE
                       & 1.2$^{+1.4}_{-0.5} \times10^{12}$ 
                       & 1.1$^{+0.9}_{-0.4} \times10^{12}$
                       & 3.6$^{+2.2}_{-1.1} \times10^{11}$
                       & 8.0$^{+5.0}_{-2.9} \times10^{11}$
                       & 3.8$^{+1.9}_{-1.2} \times10^{11}$ \\

$EM$ (\e{59}\cm{-3})   & 1.2$^{+0.3}_{-0.1}$ 
                       & 1.2$^{+0.2}_{-0.2}$ 
                       & 0.6$^{+0.1}_{-0.06}$ 
                       & 0.8$^{+0.1}_{-0.1}$
                       & 0.9$^{+0.1}_{-0.06}$
                       & 0.9$^{+0.1}_{-0.1}$ \\

$\delta \chi^2$/dof  & 1.12/134 & 1.02/130 & 1.19/137  & 1.27/143 
                     & 0.88/143 & 0.97/145 \\
\enddata
\end{deluxetable}


\begin{deluxetable}{lccc}
\tabletypesize{\scriptsize}
\tablecaption{Energy ranges of narrow-band images used for deriving
continuum-subtracted and equivalent-width images in \candy.\label{tab-ew}}
\tablewidth{0pt}
\tablehead{
\colhead{Emission Line}  & \colhead{Line} & \colhead{Left Continuum}
 & \colhead{Right Continuum} \\
\colhead{ }  & \colhead{ev} & \colhead{ev} & \colhead{ev} 
}
\startdata
Si & 1750--1950 & 1550--1700 & 2100--2250 \\
S  & 2350--2600 & 2100--2250 & 2600--2800 \\
\enddata

\end{deluxetable}


\clearpage

\clearpage

\clearpage

\clearpage

\end{document}